\pdfoutput=1
\documentclass[fleqn,usenatbib]{mnras}

\usepackage{newtxtext,newtxmath}

\usepackage[T1]{fontenc}
\usepackage{ae,aecompl}

\usepackage{graphicx}	
\usepackage{amsmath}	
\usepackage{float}
\usepackage{gensymb}
\usepackage{tabu}
\usepackage{animate}
\usepackage{media9}
\usepackage{enumerate}
\title[When is morphology imprinted on galaxies?]{\textit{SDSS-IV MaNGA}: When is morphology imprinted on galaxies?}

\author[T. Peterken et al.]{
Thomas Peterken,$^{1}$\thanks{E-mail: Thomas.Peterken@nottingham.ac.uk}
Michael Merrifield,$^{1}$\thanks{E-mail: Michael.Merrifield@nottingham.ac.uk}
Alfonso Arag{\'o}n-Salamanca,$^{1}$
\newauthor
Vladimir Avila-Reese,$^{2}$ Nicholas F.\ Boardman,$^{3}$ Niv Drory,$^{4}$ Richard R. Lane$^{5}$\\
$^{1}$School of Physics and Astronomy, University of Nottingham, University Park, Nottingham NG7 2RD, UK\\
$^{2}$Instituto de Astronom{\'i}a, Universidad Nacional Aut{\'o}noma de M{\'e}xico, A.P. 70--264, 04510 CDMX, M{\'e}xico\\
$^{3}$Department of Physics and Astronomy, University of Utah, Salt Lake City, UT 84112, USA\\
$^{4}$McDonald Observatory, The University of Texas at Austin, 1 University Station, Austin, TX 78712, USA\\
$^{5}$Instituto de Astronom{\'i}a y Ciencias Planetarias de Atacama, Universidad de Atacama, Copayapu 485, Copiap{\'o}, Chile
}

\date{Draft copy: \today}

\pubyear{2020}

\begin{document}
\label{firstpage}
\pagerange{\pageref{firstpage}--\pageref{lastpage}}
\maketitle

\begin{abstract}
It remains an open question as to how long ago the morphology that we see in a present-day galaxy was typically imprinted. Studies of galaxy populations at different redshifts reveal that the balance of morphologies has changed over time, but such snapshots cannot uncover the typical timescales over which individual galaxies undergo morphological transformation, nor which are the progenitors of today's galaxies of different types.  However, these studies also show a strong link between morphology and star-formation rate over a large range in redshift, which offers an alternative probe of morphological transformation.  We therefore derive the evolution in star-formation rate and stellar~mass of a sample of 4342 galaxies in the SDSS-IV MaNGA survey through a stellar~population ``fossil~record'' approach, and show that the average evolution of the population shows good agreement with known behaviour from previous studies.  Although the correlation between a galaxy's contemporaneous morphology and star-formation rate is strong over a large range of lookback times, we find that a galaxy's present-day morphology only correlates with its relatively recent ($\sim 2\,\textrm{Gyr}$) star-formation history.  We therefore find strong evidence that morphological transitions to galaxies' current appearance occurred on timescales as short as a few billion years.
\end{abstract}




\section{Introduction}
\label{sec:Intro}

Dividing a galaxy sample by morphology was the earliest classification scheme for these objects \citep{Hubble26neb, Hubble36}, and morphology is still considered to be one of the defining characteristics of a galaxy, which is closely tied to its other properties.  For example, a galaxy's morphology is strongly correlated with its colour (e.g.\ \citealt{Strateva+01, Baldry+04}) and hence its star-formation rate (SFR; e.g.\ \citealt{Brinchmann+04, Noeske+07, CanoDiaz+19}), with late-type or spiral galaxies generally lying along the star-forming ``main~sequence'' in the stellar mass--SFR plane.  The population of galaxies with lower star-formation rates than the ``main sequence'' has grown over time \citep{Peng+10} to constitute the bulk of the stellar mass in the present-day Universe \citep{Salim+07}, and comprises mainly early-type galaxies.  The two galaxy populations are distinct, and are separated by an underpopulated region known as the ``green~valley'' (see e.g.\ \citealt{Schawinski+14}).  Studies of galaxy populations in the nearby Universe have revealed that this structure has been in place at least since $z\sim0.5$, with galaxies of different morphologies consistently populating distinct parts of stellar mass--SFR plane \citep{Kauffmann+03, Moresco+13, CanoDiaz+16, Wang+20}.  At higher redshifts this picture breaks down somewhat, but there is some evidence linking morphological structure and star-formation even as early as $z\sim2.5$ \citep{Wuyts+11, Bell+12, Mortlock+13, Lang+14}.

From redshift snapshot studies, it is not possible to measure how rapidly properties are able to change in a single galaxy.  The exact role of a galaxy's morphology as it transitions from the star-forming to the retired population is therefore still unclear (see e.g.\ \citealt{Martig+09, Cheung+12, LopezFernandez+18, CanoDiaz+19, Cluver+20}). By analysing the spatially-resolved stellar populations contained within galaxies, it is possible to start to understand their evolution on a galaxy-by-galaxy basis.  We previously used such a ``fossil~record'' approach to investigate the morphological evolution of spiral galaxies (\citealt{Peterken+20}; see also \citealt{IbarraMedel+16}).  However, these ``time~slicing'' analyses are unable to account for the rearranging of stars within galaxies through mergers and secular evolution (see e.g.\ \citealt{Kormendy13}), or for the fading of spiral structure in older populations \citep{Peterken+19TS}.  Fortunately, there is an alternative complementary approach that we can adopt: the star-formation history (SFH) of an entire galaxy can be used to trace its past.  \citet{LopezFernandez+18} and \citet{Sanchez+19} have recently shown that it is possible to derive the distribution of galaxies in the $M_{\star}$--SFR plane at different lookback times using a fossil~record approach.  Here, we combine a new implementation of such fossil~record analysis for galaxies in the SDSS-IV~MaNGA survey \citep{Bundy+15} with citizen-science-derived data on their present-day morphologies.  By comparing galaxies' morphological histories associated with their past star formation to their observed present-day morphologies, we are able to estimate when their current forms were imprinted.


Throughout this paper, we assume a flat~$\Lambda$CDM cosmology with $H_0=68\,\textrm{km}\,\textrm{s}^{-1}\,\textrm{Mpc}^{-1}$ and $\Omega_{\rm m}=0.308$, consistent with \citet{Planck15}.

\section{MaNGA}
\label{sec:Data-MaNGA}

Integral~field spectroscopic (IFS) galaxy surveys offer an ideal tool for studying the history of low~redshift galaxies.  Fossil~record analysis allows for the evolution of a single galaxy population to be studied, thereby ensuring that low-mass galaxies are included at all lookback times and avoiding redshift-dependent sampling effects such as progenitor bias.  It is also possible to measure the entire stellar population of each galaxy to a consistent physical radius, independent  of the galaxy's distance.

With its large sample size, Mapping Nearby Galaxies at Apache Point Observatory (MaNGA; \citealt{Bundy+15}) is therefore an ideal survey for a fossil~record analysis.  MaNGA is a large IFS survey and is part of the fourth generation of the Sloan Digital Sky Survey (SDSS-IV; \citealt{Blanton+17}).  It uses the BOSS spectrograph \citep{Smee+13} on the 2.5-metre Sloan telescope at Apache Point Observatory \citep{Gunn+06} to obtain spectra of resolution $R\approx2000$ over the wavelength range of 3600--10300~\AA.  By the project's completion later this year, observations with a spatial resolution of 2.5~arcseconds will be acquired for 10,000 low-redshift ($0.01<z<0.15$, median $z\sim0.3$) galaxies \citep{Yan+16-design} through the use of integral-field units of five different sizes with diameters of between 12 and 32~arcsec \citep{Drory+15}.  The raw fibre spectra are calibrated to a better than 5\% accuracy \citep{Yan+16-cal} and the datacubes are produced by a dedicated data reduction pipeline (DRP; \citealt{Law+16}).  Here we also make use of some of the data analysis products made available from MaNGA's purpose-built data analysis pipeline (DAP; \citealt{DAP, DAPLines}).

Observations are based on two main subsamples which are designed to have a flat distribution in log($M_{\star}$).  The Primary sample is observed to $1.5\,R_{\rm e}$ --- where $R_{\rm e}$ is the effective radius measured by the NASA-Sloan Atlas (NSA; \citealt{NSA}) --- and the Secondary is observed to $2.5\,R_{\rm e}$.  The Primary sample is supplemented by a colour-enhanced sample --- targeting underpopulated regions of the $M_{\star}$--colour plane --- to form the combined Primary+ sample.  All observations reach a minimum signal-to-noise ratio of $5\,\textrm{\AA}^{-1}$ per fibre at $1.5\,R_{\rm e}$ in the $r$ band \citep{Law+15}.  Although the observing strategy results in samples which are not volume-limited in nature, the selection criteria are well-defined.  Weightings can therefore be calculated to convert each sample into an effectively volume-limited one for obtaining science results \citep{Wake+17, Sanchez+19}.  In this work, we use the weightings calculated by Calette et al.\ (in preparation), which are robust for stellar masses above $\sim10^{9}\,M_{\odot}$ (see \citealt{Sanchez+19}; also \citealt{RodriguezPuebla+20} for further details).

\section{Sample selection}
\label{sec:Sample}

From the latest internal MaNGA data release (MaNGA Product Launch 9; MPL-9), we selected all Primary+ galaxies.  The choice of MaNGA sample was made to ensure that all galaxies are analysed to the same relative radius whilst providing the largest sample possible.  We reiterate that the sample weightings described above ensure that the oversampling in green valley, high-mass blue, and low-mass red galaxies are appropriately down-weighted for analysis purposes.  We conservatively removed all galaxies for which the MaNGA DRP assigns any warning flags, and we also require the DAP to have run without problems for each galaxy and for which emission spectra are therefore available.  These criteria together produce a full sample of 4342 galaxies for which observations extend to $1.5\,R_{\rm e}$.

\section{Spectral fitting and star-formation history}
\label{sec:Fitting}

For each galaxy, we use the line-of-sight stellar velocity measurements from the DAP to rebin each spaxel's spectrum onto a common sampling of rest wavelengths.  We then subtract the DAP's model of each spaxel's emission spectrum from the observed spectra and sum all spectra from within $1.2\,R_{\rm e}$ to obtain a single spectrum of the stellar component for each galaxy.  This aperture was chosen to balance the inclusion of as much data as possible while avoiding overlap with the hexagonal MaNGA IFU edges which might bias results (see e.g.\ \citealt{IbarraMedel+16}).

We then applied the same spectral fitting methods detailed by \citet{Peterken+20} to each of the 4342 emission-subtracted rest-frame galaxy spectra.  The fitting method uses \textsc{Starlight} \citep{Starlight} to find a best-fit combination of 54 single stellar population (SSP) templates from E-MILES (\citealt{E-MILES}; covering ages between $10^{7.85}$ and $10^{10.25}\,\textrm{years}$ and metallicities in the range $-1.71\leq[{\rm M}/{\rm H}]\leq+0.22$) and twelve from \citeauthor{Asa'd+17} (\citeyear{Asa'd+17}; with ages between $10^{6.8}$ and $10^{7.6}\,\textrm{years}$ and metallicities of $[\textrm{M}/\textrm{H}] = -0.41, +0.00$).  The SSP spectra all assume ``Padova'' \citep{Bertelli+94, Girardi+00} isochrones, a \citet{Chabrier03} IMF, and Milky-Way metallicity-scaled $\alpha$-element enhancement (``baseFe'').  We also allow for a single dust extinction following a \citet{Calzetti+00} law, and fit within the range $3541.4\leq\lambda\leq8950.4\,\textrm{\AA}$.  We use \textsc{Starlight} in a ``long fit'' configuration to prioritise robustness over speed, which we demonstrated in \citet[see Appendix~A]{Peterken+20} results in reliable derived SFHs for stellar populations older than 30~Myr.  We refer the reader to \citet{Peterken+20} for full details of the fitting method.

From the distribution of template weights assigned by \textsc{Starlight} to each SSP template in its best-fit model, it is possible to derive a SFH for each galaxy.  We use the SSP initial (i.e.\ formation) mass weights, which utilise the mass-to-light ratios of the E-MILES spectra and the mass-loss estimates for each template.  We then measure a ``raw'' SFH by assigning each SSP a temporal bin for which the SSP's nominal age lies in the centre of that bin in log(age).\footnote{In the case of the oldest template, the upper box boundary is chosen to satisfy this centring criterion, resulting in the raw SFHs being sampled at lookback times older than the age of the Universe.  For the youngest templates, the lower age-bound of the box is taken to be $0\,\textrm{Gyr}$.} Each SSP's total mass contribution to the spectrum is assumed to be distributed evenly over its respective bin, so that the raw SFH measured in $M_{\odot}$/yr is a step function when summed over all metallicities, which we sample at 250 log-spaced stellar population ages.  These raw step-function SFHs are smoothed using a Gaussian of width 0.2~dex in age to produce a realistically-smooth SFH for each galaxy; the exact smoothing adopted has no effect on any of the conclusions that we draw.

\section{Evolution of the mass function and the ``main~sequence'' for star~formation}
\label{sec:MS}

\begin{figure}
    \centering
    \animategraphics[width=\columnwidth, autoplay, loop, controls]{12.5}{animations/MainSequence/SFRt/}{0000}{0115}
    \caption{The evolution of the star-formation ``main~sequence'', with galaxies coloured by their specific star-formation rate (\textit{lower~left}, see text for thresholds).  Point opacities reflect the sample weighting of each galaxy.  \textit{Top}: The evolution of the (appropriately sample-weighted) galaxy stellar~mass function and its contributions from galaxies of different specific star-formation rates.  \textit{Right}: The distributions in star-formation rate of galaxies in each specific star-formation rate classification.}
    \label{anim:MS_sfr}
\end{figure}

From the smoothed SFHs from the \textsc{Starlight} fits, we are able to determine the instantaneous star-formation rate of each galaxy at any lookback time.  By considering the cumulative sum of mass weights older than a specific lookback time, we are also able to determine the stellar mass $M_{\star}$ of each galaxy at that time.  We can therefore derive how the $M_{\star}$--SFR plane is populated at any redshift, as demonstrated by \citet{LopezFernandez+18} and \citet{Sanchez+19}.  To do so, we first register each galaxy's SFH to account for its observed redshift, which imposes a lower limit of $10^{8.83}\,\textrm{years}$ ($0.68\,\textrm{Gyr}$) in lookback time, younger than which we cannot measure due to loss of sample completeness.

We show the derived ``main~sequence'' of star formation between lookback times of 10 and 0.68~Gyr in Figure~\ref{anim:MS_sfr}. The effect of downsizing can be readily seen, in that high-mass galaxies began to exhibit significantly declining star formation around $5-7\,\textrm{Gyr}$ ago, while the galaxies at the low-mass end of the sample only began to decline in star-formation approximately $1-2\,\textrm{Gyr}$ ago.  This effect is in agreement with other studies (e.g.\ \citealt{Muzzin+13, Sanchez+19}) and will be explored in further detail and quantified in a forthcoming paper (Peterken et al.\ in preparation).  One can also see growth in the number density of high-mass ($\geq10^{11}\,M_{\odot}$) galaxies as the population builds in mass over time, although the mass~function evolution since $z\sim1$ in high-mass galaxies is small, confirming similar results found by cosmological studies (e.g.\ \citealt{Muzzin+13, Wright+18, Leja+20}).

\begin{figure}
    \centering
    \animategraphics[width=\columnwidth, autoplay, loop, controls]{12.5}{animations/MainSequence/GZ/}{0000}{0115}
    \caption{As Figure~\ref{anim:MS_sfr}, but with galaxies coloured by their present-day morphological classifications.}
    \label{anim:MS_gz}
\end{figure}

\subsection{Specific star-formation rate effects}
\label{sec:MS-SFR}

In Figure~\ref{anim:MS_sfr}, galaxies are coloured according to their contemporaneous specific SFR:
\begin{itemize}
    \item Star-forming: $\textrm{sSFR}_{t} > \frac{0.2}{A_t}\,\textrm{yr}^{-1}$
    \item Retiring: $\frac{0.02}{A_t} < \textrm{sSFR}_{t} \leq \frac{0.2}{A_t}\,\textrm{yr}^{-1}$
    \item Retired: $\textrm{sSFR}_{t} \leq \frac{0.02}{A_t}\,\textrm{yr}^{-1}$
\end{itemize}
where $\textrm{sSFR}_{t}$ is the ratio of each galaxy's SFR at lookback time $t$ to its stellar mass $M_{\star}$ which had built up by that time and $A_t$ is the Universe's age at $t$.  The threshold between retiring and star-forming galaxies is therefore the same as that used by \citet{Pacifici+16} and \citet{Sanchez+19} to separate retired and star-forming populations.

Although we do not find significant evolution in the overall stellar mass function, we are able to measure how the contributions to the mass function from galaxies of different specific SFRs have varied over time. We see that the mass functions of retired and retiring galaxies have built rapidly over the last 7 and $5\,\textrm{Gyr}$ respectively at the high-mass end of the sample (at the expense of that of the star-forming sample) before a more modest growth at the low-mass end of the sample.  By the present day, we find that retired galaxies constitute the majority of the galaxy sample above $\sim2\times10^{11}\,M_{\odot}$, but that the galaxy stellar mass function of the whole population is still dominated by star-forming galaxies with $M_{\star}<10^{10}\,M_{\odot}$ at $t=0.68\,\textrm{Gyr}$.


As the ``star forming'', ``retiring'', and ``retired'' classifications are broadly analogous to colour-based ``blue cloud'', ``green valley'', and ``red sequence'' designations respectively (as we will explore in more detail in Peterken et al.\ in preparation), the results found here imply that the red sequence has increased its contribution to the high-mass end of the total mass function over the last $5\,\textrm{Gyr}$.  These results are in good agreement with with the equivalent findings from studies of galaxy populations at different redshifts (e.g.\ \citealt{Bell+12, Muzzin+13, Ilbert+13, Davidzon+17}), offering some assurance as to the reliability of  star-formation histories derived through this fossil~record.

\subsection{Morphological effects}
\label{sec:MS-Morph}

Of the 4342 galaxies used here, 3969 (91.4\%) have been classified by the Galaxy Zoo project \citep{Lintott+08}, in which volunteer ``citizen~scientists'' are asked to identify galaxies' morphological features.  We make use of data from Galaxy~Zoo~2 \citep{Willett+13} using the redshift-debiased vote fractions of \citet{Hart+16} to split the galaxy sample into broad present-day morphology classes.  We wish to minimise the number of ``ambiguous'' morphologies, so we adopt less conservative thresholds for classifications than we previously used to select spiral galaxies in \citet{Peterken+20}.  Specifically, we classify galaxies according to the following criteria:
\begin{itemize}
    \item Early-type: $\left(p_{\rm features\, or\, disk}<0.5\right) \cup (p_{\rm spiral}<0.5)$
    \item Late-type: $\left(p_{\rm features\, or\, disk}>0.5\right) \cap (p_{\rm spiral}>0.5)$
\end{itemize}
where $p_{\rm [class]}$ indicates the redshift-debiased vote fractions from \citet{Hart+16}.

Figure~\ref{anim:MS_gz} shows the evolution of galaxies in the $M_{\star}$--SFR plane coloured by present-day morphology.  In agreement with other studies (e.g.\ \citealt{RodriguezPuebla+20}), we find that the galaxy stellar mass functions of early- and late-type morphologies are different at $z=0$, with early-type galaxies dominating the high-mass end of the sample.  Over most of the range in stellar mass of our sample, the population of early-type galaxies here will be dominated by fast-rotator galaxies (see e.g.\ \citealt{Bell+12,Cappellari16,Wang+20}), but we find that the progenitors of the most massive present-day early-type galaxies --- likely `true' slow-rotator ellipticals --- have \textit{always} had high stellar mass compared to the galaxy population as a whole, reflecting the minimal change in rank of galaxy masses: massive galaxies have always been massive, as we previously found in spirals \citep{Peterken+20}.

However, despite the variation in their stellar mass distributions, we find that galaxies of all ultimate morphologies are well mixed at early times in the $M_{\star}$--SFR plane (as previously found by \citealt{LopezFernandez+18}) and remain so until the last $\sim2\,\textrm{Gyr}$ (i.e.\ since $z\sim0.16$), when they separate out into the expected segregated regions. Since location in this plane correlates strongly with morphology over longer cosmic timescales than this \citep{Wuyts+11, Cheung+12, Bell+12, Lang+14}, this mixing implies that over much of its history a galaxy's morphology was not in any way dictated by its current classification, which has only been imprinted relatively recently, in the last few billion years.

\section{Further checks}
\label{sec:Limitations}

Comparing the mass functions in Figure~\ref{anim:MS_gz} to those of \citet{RodriguezPuebla+20} reveals a larger fraction of low-mass ($M_{\star}\lessapprox10^{9.5}\,M_{\odot}$) galaxies classified as early-type in the Galaxy Zoo analysis. This difference likely arises from the difficulties inherent in visually classifying these smaller systems using ground-based images of limited resolution.  We will explore this issue further in Peterken et al. (in preparation).  However, note that it does not compromise the results found here, since the disconnect between current morphology and SFH is apparent across all masses.  As a further check for any dependence on the method used to classify galaxy morphology, we repeated the analysis presented here with the machine~learning classifications of \citet{DominguezSanchez+18} for galaxies in the SDSS public Data~Release~15 and found the same result of a clear separation only occurring within the last $\sim2\,\textrm{Gyr}$.

One further issue might be that, with the galaxies only being analyzed out to a radius of $1.2\,R_{\rm e}$, we might be introducing some bias due to the known radial variations in SFH (see e.g.\ \citealt{IbarraMedel+16, Peterken+20}).  As a check, we repeated the analysis of Section~\ref{sec:MS-Morph} using MaNGA's Secondary sample, for which we are able to measure SFHs out to $2.3\,R_{\rm e}$, and found that the results are unchanged with an aperture almost twice as large.

\section{Discussion and conclusions}
\label{sec:Conclusions}

We have used spectral fitting methods with \textsc{Starlight} to measure star-formation histories of the inner $1.2\,R_{\rm e}$ of a sample of 4342 galaxies, which when appropriately weighted form an effectively volume-complete sample for present-day stellar masses $M_{\star}>10^{9}\,M_{\odot}$.  We derived the positions of these galaxies in the stellar~mass--star-formation~rate plane at many lookback times between 10 and $0.68\,\textrm{Gyr}$ to illustrate the evolution of the ``main sequence'' of star formation and the galaxy mass function.  We showed that such an approach is able to recover the known downsizing effects found by studies of galaxy populations at different redshifts (e.g.\ \citealt{Peng+10}) and in other fossil~record analyses (e.g.\ \citealt{Sanchez+19}), in that the galaxies with highest present-day stellar mass exhibited declining star~formation at earlier times than low-mass galaxies, causing the mass function of retired and retiring (or equivalently red~sequence) galaxies to grow rapidly at the high-mass end, starting $5\,\textrm{Gyr}$ ago.  This result has previously been seen in studies of galaxy populations (e.g.\ \citealt{Muzzin+13, Ilbert+13, Davidzon+17}), but is recovered here using an entirely independent and complementary approach, providing reassurance as to the efficacy of this approach.

By splitting the sample into sub-samples of different morphologies using the Galaxy~Zoo classifications, we found that the regions of the $M_{\star}$--SFR plane inhabited by present-day (disk-dominated fast-rotator) early- and late-type galaxies have only been systematically different for approximately the last $2\,\textrm{Gyr}$.  Since location in this plane is, in itself, a good proxy for morphology at greater lookback times than this, we conclude that a galaxy's current form is not connected to its historical morphology.  It is therefore apparent that a galaxy's current morphology has only been established in the last few billion years.

We note here that this integrated spectral approach cannot distinguish between secular galaxy growth and the effects of mergers, and the latter may become a significant factor for the higher-mass early-type galaxies studied here (e.g.\ \citealt{NaabBurkert03, Bournaud+07}).  Mergers will impact to some extent on the mass-function results, since our detection of a single galaxy may in reality be a combination of two or more lower-mass galaxies comprising its pre-merger progenitor galaxies.  However, merger effects will serve only to randomise structure and further shorten the timescale over which the current morphology has been in place, and therefore do not compromise the main conclusion about the relatively recent imprint of morphology.

While a galaxy's stellar mass is found to tell us something about its past history --- a result also found by others (e.g.\ \citealt{IbarraMedel+16, GarciaBenito+19}) --- its current morphology cares very little about its more distant morphological past.

\section{Data Availability}

This publication uses the team-internal MPL-9 version of the MaNGA science data products, which are broadly similar to previous versions available publicly through SDSS Data Release DR15 \citep{DR15} and other earlier releases.  Comparable data products containing the full MaNGA galaxy sample --- including the full sample used here --- will be publicly released in 2021 as part of SDSS DR17, as will the raw data and all previous versions of the data reduction pipeline.

\section*{Acknowledgements}

The authors thank S.\ F.\ S{\'a}nchez, D.\ Wake, A.\ R.\ Calette and A.\ Rodriguez-Puebla for their extensive help and support on the technical aspects of this work.

Funding for the Sloan Digital Sky Survey IV has been provided by the Alfred P. Sloan Foundation, the U.S. Department of Energy Office of Science, and the Participating Institutions. SDSS acknowledges support and resources from the Center for High-Performance Computing at the University of Utah. The SDSS web site is \url{www.sdss.org}.

SDSS is managed by the Astrophysical Research Consortium for the Participating Institutions of the SDSS Collaboration including the Brazilian Participation Group, the Carnegie Institution for Science, Carnegie Mellon University, the Chilean Participation Group, the French Participation Group, Harvard-Smithsonian Center for Astrophysics, Instituto de Astrof{\'i}sica de Canarias, The Johns Hopkins University, Kavli Institute for the Physics and Mathematics of the Universe (IPMU) / University of Tokyo, the Korean Participation Group, Lawrence Berkeley National Laboratory, Leibniz Institut f{\:u}r Astrophysik Potsdam (AIP), Max-Planck-Institut f{\:u}r Astronomie (MPIA Heidelberg), Max-Planck-Institut f{\:u}r Astrophysik (MPA Garching), Max-Planck-Institut f{\:u}r Extraterrestrische Physik (MPE), National Astronomical Observatories of China, New Mexico State University, New York University, University of Notre Dame, Observat{\'o}rio Nacional / MCTI, The Ohio State University, Pennsylvania State University, Shanghai Astronomical Observatory, United Kingdom Participation Group, Universidad Nacional Aut{\'o}noma de M{\'e}xico, University of Arizona, University of Colorado Boulder, University of Oxford, University of Portsmouth, University of Utah, University of Virginia, University of Washington, University of Wisconsin, Vanderbilt University, and Yale University.

This publication uses data generated via the Zooniverse.org platform, development of which is funded by generous support, including a Global Impact Award from Google, and by a grant from the Alfred P. Sloan Foundation.

We are grateful for access to the University of Nottingham's \texttt{Augusta} high-performance computing facility.




\bibliographystyle{mnras}
\bibliography{refs} 




\label{lastpage}
\end{document}